\documentclass[12pt]{cernart}
\usepackage{amsmath}
\usepackage{epsfig}

\newcommand{\piplus}   {{\pi}^{+}}
\newcommand{\piminus}  {{\pi}^{-}}
\newcommand{\pizero}   {{\pi}^{0}}

\newcommand{\Kzero}    {{\mathrm K}_{\phantom{S}}^{0}}
\newcommand{\Kzerobar} {{\overline{\mathrm K}}_{\phantom{S}}^{0}}
\newcommand{\KS}       {{\mathrm K}_{\mathrm S}}
\newcommand{\KL}       {{\mathrm K}_{\mathrm L}}

\newcommand{\epsprime}  {\epsilon^{\,\prime}}
\newcommand{\Reee}      {\Real(\epsprime/\epsilon)}

\newcommand{\alphaLS}   {\alpha_{\mathrm{LS}}}
\newcommand{\alphaSL}   {\alpha_{\mathrm{SL}}}

\newcommand{\tauS}      {\tau_{\mathrm{S}}}
\newcommand{\tauL}      {\tau_{\mathrm{L}}}
\newcommand{\NS}        {N_{\mathrm{S}}}
\newcommand{\NL}        {N_{\mathrm{L}}}

\newcommand{\tS}        {t_{\mathrm{S}}}
\newcommand{\tL}        {t_{\mathrm{L}}}
\newcommand{\tSL}       {t_{\mathrm{S,L}}}
\newcommand{\mK}        {m_{\mathrm{K}}}
\newcommand{\mS}        {m_{\mathrm{S}}}
\newcommand{\mL}        {m_{\mathrm{L}}}
\newcommand{\pK}        {p_{\mathrm{K}}}
\newcommand{\EK}        {E_{\mathrm{K}}}

\newcommand{\ctau}      {\tau}

\newcommand{\zAKS}      {z_{\mathrm{AKS}}}

\newcommand{\ztargSL}   {z_{\mathrm{targ\,S,L}}}
\newcommand{\DE}        {D(\EK)}
\newcommand{\DS}        {D_{\mathrm{S}}}
\newcommand{\DL}        {D_{\mathrm{L}}}
\newcommand{\DSL}       {D_{\mathrm{S,L}}}
\newcommand{\fS}        {f_{\mathrm{S}}}
\newcommand{\fL}        {f_{\mathrm{L}}}
\newcommand{\fSL}       {f_{\mathrm{S,L}}}

\newcommand{\proton}    {\mathrm{p}}
\newcommand{\pippim}    {\piplus\piminus}
\newcommand{\pizpiz}    {\pizero\pizero}
\newcommand{\zzz}       {\pizero\pizero\pizero}
\newcommand{\Kzpipi}    {\Kzero\to\pi\pi}
\newcommand{\Kzpm}      {\Kzero\to\piplus\piminus}
\newcommand{\Kzzz}      {\Kzero\to\pizero\pizero}
\newcommand{\KSpipi}    {\KS\to\pi\pi}
\newcommand{\KLpipi}    {\KL\to\pi\pi}
\newcommand{\KSpm}      {\KS\to\piplus\piminus}
\newcommand{\KLpm}      {\KL\to\piplus\piminus}
\newcommand{\KSzz}      {\KS\to\pizero\pizero}
\newcommand{\KLzz}      {\KL\to\pizero\pizero}
\newcommand{\KLzzz}     {\KL\to\pizero\pizero\pizero}
\newcommand{\Kethree}   {\mathrm{K}_{\mathrm{e3}}}
\newcommand{\Kmuthree}  {\mathrm{K}_{\mu 3}}

\newcommand{\ptp}       {p_{\mathrm{T}}^{\,\prime}}
\newcommand{\ptpsq}     {p_{\mathrm{T}}^{\,\prime\,2}}

\newcommand{\rcog}      {C_{\mathrm{g}}}
\newcommand{\lamext}    {\lambda_{\mathrm{ext}}}
\newcommand{\Acoll}     {A_{\mathrm{coll}}}
\newcommand{\tcoll}     {t_{\mathrm{coll}}}
\newcommand{\zcoll}     {z_{\mathrm{coll}}}
\newcommand{\Real}      {\mathrm{Re}}
\newcommand{\E}[1]      {\times 10^{#1}}
\newcommand{\degree}    {^\circ}

\newcommand{\EPJ}  [3] {E. Phys. J. {\bf C#1} (#2) #3}
\newcommand{\NIM}  [3] {Nucl. Inst. Meth. {\bf A#1} (#2) #3}
\newcommand{\PL}   [3] {Phys. Lett. {\bf B#1} (#2) #3}
\newcommand{\PRL}  [3] {Phys. Rev. Lett. {\bf #1} (#2) #3}
\newcommand{\ZFP}  [3] {Z. Phys. {\bf C#1} (#2) #3}
\newcommand{\etal}    {{\it{et~al}.}}

\begin{document}

\begin{titlepage}
\docnum{CERN--EP/2002--028}
\date{26 April 2002}
\title{
  \Large \bf A Measurement of the $\mathbf{K_S}$ Lifetime}

\begin{Authlist}

{\bf The NA48 Collaboration} \\

\vspace*{10mm}

\begin{center}
A.~Lai,
 D.~Marras \\
{\em Dipartimento di Fisica dell'Universit\`a e Sezione dell'INFN di Cagliari, I-09100 Cagliari, Italy.} \\[0.2cm]
J.R.~Batley,
A.~Bevan\footnotemark[1],
 R.S.~Dosanjh,
 T.J.~Gershon\footnotemark[2],
B.~Hay\footnotemark[3],
G.E.~Kalmus,
 C.~Lazzeroni,
 D.J.~Munday,
M.D.~Needham\footnotemark[4],
E.~Olaiya,
 M.A.~Parker,
 T.O.~White,
 S.A.~Wotton \\
{\em Cavendish Laboratory, University of Cambridge, Cambridge, CB3 0HE, U.K.\footnotemark[5].} \\[0.2cm] 
G.~Barr,
 G.~Bocquet,
 A.~Ceccucci,
 T.~Cuhadar-D\"{o}nszelmann,
 D.~Cundy,
 G.~D'Agostini,
 N.~Doble,
V.~Falaleev,
W.~Funk,
 L.~Gatignon,
 A.~Gonidec,
 B.~Gorini,
 G.~Govi,
 P.~Grafstr\"om,
W.~Kubischta,
 A.~Lacourt,
M.~Lenti\footnotemark[6],
S.~Luitz\footnotemark[7],
J.P.~Matheys,
 I.~Mikulec\footnotemark[8],
A.~Norton,
 S.~Palestini,
 B.~Panzer-Steindel,
D.~Schinzel,
G.~Tatishvili\footnotemark[9],
H.~Taureg,
 M.~Velasco,
O.~Vossnack,
 H.~Wahl \\
{\em CERN, CH-1211 Gen\`eve 23, Switzerland.} \\[0.2cm] 
C.~Cheshkov,
A.~Gaponenko\footnotemark[10],
 P.~Hristov,
V.~Kekelidze,
 D.~Madigojine,
N.~Molokanova,
Yu.~Potrebenikov,
 A.~Tkatchev,
 A.~Zinchenko \\
{\em Joint Institute for Nuclear Research, Dubna, Russian    Federation.} \\[0.2cm] 
I.~Knowles,
 V.~Martin,
H.~Parsons,
 R.~Sacco,
 A.~Walker \\
{\em Department of Physics and Astronomy, University of    Edinburgh, Edinburgh,    EH9 3JZ, U.K.\footnotemark[5].} \\[0.2cm] 
M.~Contalbrigo,
 P.~Dalpiaz,
 J.~Duclos,
P.L.~Frabetti\footnotemark[11],
 A.~Gianoli,
 M.~Martini,
 F.~Petrucci,
 M.~Savri\'e,
M.~Scarpa \\
{\em Dipartimento di Fisica dell'Universit\`a e Sezione    dell'INFN di Ferrara, I-44100 Ferrara, Italy.} \\[0.2cm] 
A.~Bizzeti\footnotemark[12],
M.~Calvetti,
 G.~Collazuol,
 G.~Graziani,
 E.~Iacopini,
 F.~Martelli\footnotemark[13],
 M.~Veltri\footnotemark[13] \\
{\em Dipartimento di Fisica dell'Universit\`a e Sezione dell'INFN di Firenze, I-50125 Firenze, Italy.} \\[0.2cm] 
H.G.~Becker,
H.~Bl\"umer,
D.~Coward,
 M.~Eppard,
 H.~Fox,
 A.~Hirstius,
 K.~Holtz,
 A.~Kalter,
 K.~Kleinknecht,
 U.~Koch,
 L.~K\"opke,
 P.~Lopes~da~Silva, 
P.~Marouelli,
 I.~Pellmann,
 A.~Peters,
S.A.~Schmidt,
  V.~Sch\"onharting,
 Y.~Schu\'e,
 R.~Wanke,
 A.~Winhart,
 M.~Wittgen \\
{\em Institut f\"ur Physik, Universit\"at Mainz, D-55099    Mainz, Germany\footnotemark[14].} \\[0.2cm] 
\newpage
J.C.~Chollet,
 S.~Cr\'ep\'e,
 L.~Fayard,
 L.~Iconomidou-Fayard,
 J.~Ocariz,
 G.~Unal,
 I.~Wingerter-Seez \\
{\em Laboratoire de l'Acc\'el\'erateur Lin\'eaire,  IN2P3-CNRS,Universit\'e de Paris-Sud, 91898 Orsay, France\footnotemark[15].} \\[0.2cm] 
G.~Anzivino,
 P.~Cenci,
 E.~Imbergamo,
 P.~Lubrano,
 A.~Mestvirishvili,
 A.~Nappi,
M.~Pepe,
 M.~Piccini \\
{\em Dipartimento di Fisica dell'Universit\`a e Sezione    dell'INFN di Perugia, I-06100 Perugia, Italy.} \\[0.2cm] 
L.~Bertanza,
A.~Bigi,
P.~Calafiura,
R.~Carosi, 
R.~Casali,
 C.~Cerri,
 M.~Cirilli\footnotemark[16],
F.~Costantini,
 R.~Fantechi,
 S.~Giudici,
 I.~Mannelli, 
G.~Pierazzini,
 M.~Sozzi \\
{\em Dipartimento di Fisica, Scuola Normale Superiore e Sezione     dell'INFN di Pisa, I-56100 Pisa, Italy.} \\[0.2cm]   
J.B.~Cheze,
 J.~Cogan,
 M.~De Beer,
P.~Debu,
F.~Derue,
 A.~Formica,
 R.~Granier de Cassagnac,
E.~Mazzucato,
 B.~Peyaud,
 R.~Turlay,
 B.~Vallage \\
{\em DSM/DAPNIA - CEA Saclay, F-91191 Gif-sur-Yvette, France.} \\[0.2cm] 
I.~Augustin,
M.~Bender,
M.~Holder,
 A.~Maier,
 M.~Ziolkowski \\
{\em Fachbereich Physik, Universit\"at Siegen, D-57068 Siegen, Germany\footnotemark[17].} \\[0.2cm] 
R.~Arcidiacono,
 C.~Biino,
 N.~Cartiglia,
 R.~Guida,
 F.~Marchetto, 
E.~Menichetti,
 N.~Pastrone \\
{\em Dipartimento di Fisica Sperimentale dell'Universit\`a e    Sezione dell'INFN di Torino,  I-10125 Torino, Italy.} \\[0.2cm] 
J.~Nassalski,
 E.~Rondio,
 M.~Szleper,
 W.~Wislicki,
 S.~Wronka \\
{\em Soltan Institute for Nuclear Studies, Laboratory for High    Energy Physics,  PL-00-681 Warsaw, Poland\footnotemark[18].} \\[0.2cm] 
H.~Dibon,
 G.~Fischer,
 M.~Jeitler,
 M.~Markytan,
 G.~Neuhofer,
M.~Pernicka,
 A.~Taurok,
 L.~Widhalm \\
{\em \"Osterreichische Akademie der Wissenschaften, Institut  f\"ur Hochenergiephysik,  A-1050 Wien, Austria\footnotemark[19].} 
\end{center}
\end{Authlist}
\begin{center}
To be published in Physics Letters B
\end{center}
\abstract{
A measurement of the $\KS$ lifetime is presented
using data recorded by the NA48 experiment at the CERN-SPS
during 1998 and 1999.
The $\KS$ lifetime is derived from the ratio of decay time distributions
in simultaneous, collinear $\KS$ and $\KL$ beams,
giving a result which is approximately independent of the detector acceptance
and with reduced systematic errors.
The result obtained is $\tauS=(0.89598 \pm 0.00048 \pm 0.00051)\E{-10}$\,s,
where the first error is statistical and the second systematic.}
\maketitle
\footnotetext[1]{Present address: Oliver Lodge Laboratory, University of
               Liverpool, Liverpool L69 7ZE, U.K.}
\footnotetext[2]{Present address: High Energy Accelerator Research
               Organization (KEK), Tsukuba, Ibaraki, 305-0801, Japan.}
\footnotetext[3]{ Present address: EP Division, CERN, 1211 Gen\`eve 23, Switzerland.}
\footnotetext[4]{ Present address: NIKHEF, PO Box 41882, 1009 DB
  Amsterdam, The Netherlands.}
\footnotetext[5]{ Funded by the U.K.    Particle Physics and Astronomy Research Council.}
\footnotetext[6]{ On leave from Sezione dell'INFN di Firenze, I-50125
  Firenze, Italy.}
\footnotetext[7]{Present address: SLAC, Stanford, CA., 94309, USA.}
\footnotetext[8]{ On leave from \"Osterreichische Akademie der Wissenschaften, Institut  f\"ur Hochenergiephysik,  A-1050 Wien, Austria.}
\footnotetext[9]{ On leave from Joint Institute for Nuclear Research,
  Dubna, 141980, Russian Federation.}
\footnotetext[10]{Present address: University of Alberta, Edmonton Alberta T6G 2J1, Canada.}
\footnotetext[11]{Dipartimento di Fisica e INFN Bologna, viale
  Berti-Pichat 6/2, I-40127 Bologna, Italy.}
\footnotetext[12]{ Dipartimento di Fisica
dell'Universita' di Modena e Reggio Emilia, via G. Campi 213/A
I-41100, Modena, Italy.}
\footnotetext[13]{Istituto di Fisica Universit\'a di Urbino}
\footnotetext[14]{ Funded by the German Federal Minister for    Research and Technology (BMBF) under contract 7MZ18P(4)-TP2.}
\footnotetext[15]{ Funded by Institut National de Physique des
  Particules et de Physique Nucl\'eaire (IN2P3), France}
\footnotetext[16]{Present address: Dipartimento di Fisica
  dell'Universit\'a di Roma ``La Sapienza'' e Sezione INFN di Roma,
  I-00185 Roma, Italy.}
\footnotetext[17]{ Funded by the German Federal Minister for Research and Technology (BMBF) under contract 056SI74.}
\footnotetext[18]{    Supported by the Committee for Scientific Research grants 5P03B10120, 2P03B11719 and SPUB-M/CERN/P03/DZ210/2000 and using computing resources of the Interdisciplinary Center for    Mathematical and    Computational Modelling of the University of Warsaw.} 
\footnotetext[19]{    Funded by the Austrian Ministry of Education,
  Science and Culture under contract GZ 616.360/2-IV GZ
  616.363/2-VIII, and by the Fund for Promotion of Scientific Research
  in Austria (FWF) under contract P08929-PHY.}
\end{titlepage}
\renewcommand{\thefootnote}{\arabic{footnote}}

\section{Introduction}

Precise measurements of the basic physics parameters defining the neutral kaon system,
such as the masses and mean lifetimes of the $\KS$ and $\KL$ states,
are important not only in their own right
but also as essential inputs to many kaon physics analyses
such as studies of indirect and direct CP~violation or precision tests of CPT~invariance.

The $\KS$ lifetime is presently known with a relative precision of about 0.1\%~\cite{bib:pdg},
dominated by measurements
from the NA31 experiment at CERN~\cite{bib:na31}
and from the E731 and E773 experiments at Fermilab~\cite{bib:e731,bib:e773}.
Here we present a measurement of the $\KS$ lifetime
from the NA48 experiment at the CERN-SPS,
based on the same $\Kzpipi$ data samples as used for the precise determination of
the direct CP violation parameter $\Reee$~\cite{bib:na48_eprime}.
The $\KS$ lifetime is measured using an analysis technique introduced by NA31,
namely a fit to the $\KS/\KL$ ratio of decay time distributions
reconstructed in nearly collinear $\KS$ and $\KL$ beams.
This gives a result which is essentially independent
of the detector acceptance
and therefore with reduced systematic errors;\,
the $\KL$ beam is in effect used to determine the detection
efficiency for $\KS$ decays.

The analysis method is described in more detail
in the next section,
followed in Sections~\ref{sec:experiment} and~\ref{sec:selection}
by a brief summary of the main features of the NA48 experiment
and of the reconstruction and selection of $\Kzpipi$ decays;
further details can be found in~\cite{bib:na48_eprime}.
The fit used to extract the $\KS$ lifetime is described in Section~\ref{sec:results}
and the estimation of the systematic errors on the fitted lifetime
is considered in Section~\ref{sec:systematics}.

\section{The Method}                         \label{sec:method}

A defining principle of the NA48 experiment
is the simultaneous recording of decays occurring within a common decay region
traversed by two almost collinear beams of neutral kaons.
The relative target positions for each beam,
one far upstream of the decay region and the other much closer,
ensure that kaon decays in the two beams are due dominantly
to the $\KL$ or to the $\KS$ component, respectively. 
Assuming equal detection efficiencies for decays from the $\KS$ and $\KL$ beams,
the ratio $R=\NS/\NL$ of decay rates observed in each beam
as a function of the longitudinal position $z$
can be expressed as
\begin{equation}
   R(\EK,z) = A(\EK) {\fS(\tS) \over \fL(\tL)}
       \label{eqn:ratio}
\end{equation}
where $\EK$ is the kaon energy,
$A(\EK)$ is a normalisation function which
depends on the relative beam intensities
and $\tSL=(z-\ztargSL)(\mK/\pK)$
are the proper lifetimes for kaon decays in the $\KS$ or $\KL$ beams.
The functions $f(t)$ are given by
\begin{equation}
  \fSL(t) = e^{-t/\tauS} + |\eta|^2 e^{-t/\tauL}
            + 2 \DSL(\EK) |\eta| e^{-(t/\tauS + t/\tauL)/2} \cos \left( \Delta m \cdot t-\phi \right)
       \label{eqn:ft}
\end{equation}
where $\tauS$ and $\tauL$ are the $\KS$ and $\KL$ mean lifetimes,
$\Delta m=\mL-\mS$ is the mass difference between the $\KS$ and $\KL$ states,
$|\eta|$ and $\phi$ are the modulus and phase
of the ratio $A(\KLpipi)/A(\KSpipi)$ of decay amplitudes,
and the dilutions $\DSL(\EK)=[ N(\Kzero)-N(\Kzerobar) ] / [ N(\Kzero)+N(\Kzerobar) ]$
reflect the initial admixture of $\Kzero$ and $\Kzerobar$ in each beam.
In practice, in place of $z$, it is convenient to analyse the longitudinal distribution
of decay vertices in terms of the variable $\ctau = (z-\zAKS)(\mK/\pK)$
which measures the decay proper lifetime relative to the upstream edge of the decay region,
defined by the position, $\zAKS$,
of a set of scintillation counters.

Except for the $\KL$ beam at higher energies,
where the interference term contributes appreciably,
the functions $\fS(\tS)$ and $\fL(\tL)$ are dominated by the exponential terms
$e^{-\tS/\tauS}$ and $|\eta|^2 e^{-\tL/\tauL}$, respectively.
Hence the $\KS/\KL$ ratio is approximately of the form
\begin{equation*}
   R \propto {e^{-\tS/\tauS} \over e^{-\tL/\tauL}}
                \propto {e^{-\ctau (1/\tauS - 1/\tauL)}} ~.
\end{equation*}
Since $\tauL \gg\tauS$, the ratio $R$ is primarily sensitive to the $\KS$ lifetime $\tauS$.
The $\KS$ lifetime is determined by fitting a function $R(\EK,\ctau)$
of the form given in Equations~(\ref{eqn:ratio}) and~(\ref{eqn:ft})
to the ratio $\NS/\NL$
of $\Kzpipi$ decays reconstructed in the two beams.
Besides the $\KS$ lifetime itself,
the normalisation $A(\EK)$ and the dilutions $\DSL(\EK)$
are also taken as free parameters in the fit,
while the remaining physics parameters in equation~(\ref{eqn:ft}) are taken from published measurements.
Small acceptance differences between $\KS$ and $\KL$ decays
are corrected using Monte Carlo,
and background to the $\KL$ samples is subtracted using the data.
The $\pippim$ and $\pizpiz$ decay modes are analysed separately
and the results subsequently combined.

The $\KS/\KL$ ratio is reconstructed in 20 bins of energy of width $\Delta\EK=5$\,GeV
covering the range $70<\EK<170$\,GeV
and in lifetime bins of width $\Delta\ctau=0.1\tauS$.
The fit to determine the $\KS$ lifetime is carried out in the lifetime range $0.5\tauS<\ctau<3.5\tauS$.
The lower lifetime limit of $0.5\tauS$ largely avoids detector resolution effects
associated with the start of the decay region at $\ctau=0$
while the upper limit at $3.5\tauS$ is dictated by the trigger requirements.
The choice of lifetime range approximately minimises the total error on the measured $\KS$ lifetime.

\section{The NA48 experiment}                    \label{sec:experiment}

The $\KL$ and $\KS$ beams in the NA48 experiment are derived from 450\,GeV protons
incident on separate targets positioned 126\,m and 6\,m upstream
of the decay region, respectively~\cite{bib:beams}.
The $\KS$ target is located 72\,mm above the axis of the $\KL$~beam.
The $\KS$ target and collimator system
is aligned along an axis directed slightly downwards
such that the $\KS$ and $\KL$ beams subtend an angle of 0.6\,mrad
and intersect at the centre of the detector,
120\,m downstream of the $\KS$ target.

The beginning of the decay volume is accurately defined by a
$\KS$~anti-counter (AKS) located at the exit of the $\KS$~collimator.
The AKS is composed of a photon convertor followed by three scintillator counters
and is used to veto all upstream decays in the $\KS$ beam.
To minimise interactions of beam particles with air and material,
the decay region itself is contained in a 90\,m long evacuated tank
terminated by a thin composite polyamide (Kevlar) window of $3\E{-3}\,X_0$ thickness.
The tank is followed by the NA48 detector, 
the principal components of which are
a magnetic spectrometer for charged particle detection,
a liquid krypton calorimeter for photon and electron detection,
an iron-scintillator hadron calorimeter,
and muon counters
consisting of three planes of scintillator
shielded by 80\,cm thick iron walls.

The charged particle spectrometer~\cite{bib:dch} consists of two drift chambers (DCH1 and DCH2)
located before,
and two drift chambers (DCH3 and DCH4) located after,
a central dipole magnet.
Each chamber has an area of 4.5\,m$^2$
and is made up of four sets of two staggered sense wire planes
oriented along four directions
(horizontal, vertical, $\pm45^{\circ}$).
Track positions are reconstructed with a precision of 100\,$\mu$m per view,
while the momentum resolution is
$\sigma(p)/p=0.48\%\oplus 0.009 \times p$(GeV/$c$)\%.

The liquid krypton calorimeter (LKr)~\cite{bib:lkr} consists of $\sim$13000 cells,
each of cross section about 2\,cm $\times$ 2\,cm
and depth 27 radiation lengths.
For the average photon energy of 25\,GeV,
the transverse spatial resolution for photon showers is 1\,mm
and the energy resolution is 0.8\%.
The energy response is linear to about 0.1\% in the range 5$-$100\,GeV.
The photon time resolution is $\sim$500\,ps
and the $\pizpiz$ event time is known with a precision of $\sim$220\,ps.

An evacuated beam pipe of radius 8\,cm
running along the full length of the detector on its central axis
transports undecayed beam particles through each of the detector components.
The beam tube and DCH drift chambers are aligned along the bisector
of the converging $\KS$ and $\KL$ beams
in order to equalise the acceptance for $\KS$ and $\KL$ decays.
Undecayed neutral kaons in the $\KL$ beam are largely confined
within a transverse beam profile of radius $\sim 3.5$\,cm at the position of the NA48 detector.
The $\KS$ beam
has a larger beam divergence and a correspondingly larger transverse profile of about 5\,cm radius.
Due to scattering in the beam collimators,
both the $\KS$ and $\KL$ beams have an associated halo of particles extending to
larger radii from the beam axis.

The trigger for $\pizpiz$ decays~\cite{bib:neutral_trigger} requires that
the total energy deposited in the LKr calorimeter be greater than 50\,GeV,
that the kaon impact point at the calorimeter (had it not decayed)
be within 15\,cm of the beam axis,
that the decay vertex be less than 5~$\KS$~lifetimes 
from the beginning of the decay volume,
and that each of the horizontal and vertical projections
of the LKr energy distribution contain at most five peaks.
The $\pizpiz$ trigger operates with negligible deadtime
and high efficiency, $(99.920\pm0.009)$\%,
with no significant difference between $\KS$ and $\KL$ decays.
The first level of the $\pippim$ trigger~\cite{bib:charged_trigger}
is based on signals from a scintillator hodoscope
positioned in front of the LKr calorimeter and on the hit multiplicity in DCH1,
and also requires a total energy in the LKr and hadron calorimeters of at least 35\,GeV.
The second level trigger is based on tracks reconstructed using the information from
DCH1, 2 and 4,
and includes requirements that the decay vertex be less than 4.5~$\KS$~lifetimes 
from the beginning of the decay volume
and that the reconstructed mass be larger than 0.95\,$\mK$.
The efficiency of the first and second level triggers is about 99.5\% and 98.3\%,
respectively,
again with no significant difference between $\KS$ and $\KL$ decays.
The $\pippim$ trigger introduces a deadtime of about 1.1\,\%.

\section{Event Selection}                     \label{sec:selection}

$\Kzpipi$ decays are reconstructed and selected
using the same procedures and selection requirements as for the $\Reee$ analysis~\cite{bib:na48_eprime}.
The level of background remaining in the selected $\Kzpipi$ samples,
and the correction for acceptance differences between $\KS$ and $\KL$ decays,
are also evaluated using similar techniques to those in~\cite{bib:na48_eprime}.
For the determination of the $\KS$ lifetime,
it is the lifetime dependence of these corrections which is of importance,
rather than their energy dependence or their overall normalisation.

\subsection{The {\boldmath $\pi^0\pi^0$} sample}            \label{sec:selection_neutral}

The reconstruction of $\pizpiz$ events is based entirely on data from
the LKr calorimeter.
Any group of four showers,
each reconstructed within 5\,ns of their average time,
is considered.
The energy of each shower is required to lie between 3 and 100\,GeV,
and showers close to the edges of the calorimeter
(within 11\,cm of the outer edge or within 15\,cm of the central axis)
or within 2\,cm of a defective cell are excluded.
The transverse distance between any pair of showers must be greater than 10\,cm.
The position of the centre of gravity of the event
is defined as the energy-weighted average of the four shower positions.
The radial distance, $\rcog$,  of the centre of gravity from the detector axis
is required to be less than 10\,cm to suppress events due to the decay
of particles in the beam halo.

The kaon energy is estimated simply as the sum of the four shower energies.
The longitudinal decay vertex position is reconstructed from the energies and positions
of the four showers,
under the assumption that they come from the decay of a particle with
the kaon mass, $\mK$, moving along the beam axis.
A resolution of $\sim 40$-60\,cm is achieved on the vertex position,
depending on the kaon energy and decay point.
The shower pairing which best represents two $\pizero\to\gamma\gamma$ decays
is inferred using a $\chi^2$~variable
constructed from the sum, $m_1+m_2$, and difference, $m_1-m_2$,
of the two candidate $m_{\gamma\gamma}$ masses
and a parameterisation of the resolutions on $m_1\pm m_2$.

Background from $\KLzzz$ decays
is suppressed by requiring no additional showers with energy greater than 1.5\,GeV
within $\pm3$\,ns around the event time,
and by requiring $\chi^2<13.5$
(which corresponds to 3.7$\sigma$ on the $m_{\gamma\gamma}$ resolution).
All other potential sources of background to the $\pizpiz$ sample are negligibly small.

The level of $\KLzzz$ background remaining in the $\KLzz$ sample
is estimated using a control region at large values of $\chi^2$,
$36<\chi^2<135$, dominated by background.
Genuine $\KLzz$ signal events populating this control region
are first subtracted using the $\KSzz$ data, for which background is negligible.
Small differences in the shape of the $\chi^2$ distribution for $\KSzz$ and $\KLzz$ decays
are taken into account 
by applying a correction derived from Monte Carlo.
The $\KLzzz$ background in the signal region, $\chi^2<13.5$,
is estimated by extrapolating uniformly from the control region,
with an additional factor $\lamext=1.2\pm0.2$ derived from Monte Carlo simulation.
The background estimation is carried out separately for
each bin of kaon energy and proper lifetime.
At low energy and at low lifetime,
the fraction of $\zzz$ background in the $\pizpiz$ sample is negligibly small,
but rises with both energy and lifetime,
reaching about 3\,\% for $\EK=170$\,GeV and $\ctau=3.5\tauS$.
An increase of the background fraction with lifetime is to be expected
since, when only four photons are detected in the LKr calorimeter,
the reconstructed vertex position for a $\KLzzz$ decay
is shifted downstream from its true position,
while $\KLzzz$ decays occuring upstream of the decay region
are largely removed by the $\KL$ beam collimators.

\subsection{The {\boldmath $\pi^+\pi^-$} sample}           \label{sec:selection_charged}

$\Kzpm$ events are reconstructed from oppositely charged pairs of tracks
found in the magnetic spectrometer.
Each track is required to have momentum greater than 10\,GeV
and to lie at least 12\,cm from the centre of each DCH.
Each track is also required to lie within the acceptance of the LKr calorimeter
and the muon counters after extrapolation downstream.
The momentum-weighted average of the track positions
after extrapolation to the LKr calorimeter is used to define
the position of the centre of gravity of the event.
Its radial distance, $\rcog$, from the detector axis
is required to be less than 10\,cm.

The separation between the two tracks at their point of closest approach
after extrapolation upstream from the spectrometer
is required to be less than 3\,cm.
The point midway between the tracks at their closest approach
is used to define the decay vertex position.
The resolution on the longitudinal position of the decay vertex is in the range $\sim 30$-50\,cm,
while the transverse resolution is about 2\,mm.
The kaon momentum is computed from the opening angle $\theta$ of the two tracks
and from the ratio $p_+ / p_-$ of their momenta,
assuming that the event corresponds to a $\Kzpm$ decay.
Thus the reconstruction of both the decay vertex position and the kaon momentum,
and hence also of the proper time for the decay,
depend only on the geometry of the detector.

Background to the $\pippim$ sample from $\Lambda\to\proton\piminus$ decays
is reduced to a negligible level by applying an energy dependent upper cut
on the track momentum asymmetry
$|p_+-p_-|/(p_++p_-)$.
This cut also serves to remove events with tracks at low radius to the beam axis
in a way which depends only on the momentum ratio of the two tracks.
Background from $\KL\to\pi$e$\nu$ ($\Kethree$) decays 
is suppressed by requiring $E/p$ to be less than 0.8 for each track,
while $\KL\to\pi\mu\nu$ ($\Kmuthree$) decays
are suppressed by rejecting events where one or both of the tracks
is associated with a signal in the muon counters within $\pm 4$\,ns.
Additional suppression of both $\Kethree$ and $\Kmuthree$ decays is obtained
by requiring that the $\pippim$ invariant mass, $m_{\pi\pi}$, be compatible with the kaon mass
to within 3$\sigma$ of the (energy-dependent) mass resolution,
and by requiring a small missing transverse momentum,
$\ptpsq<200$\,MeV$^2$/$c^2$,
where $\ptp$ is the component of the kaon momentum
perpendicular to the line joining the production target
(identified from the vertical position of the decay vertex,
as described below)
and the point where the kaon trajectory crosses the plane of the first drift chamber.
The quantity $\ptp$ has approximately the same resolution for $\pippim$ decays
in both the $\KS$ and $\KL$ beams.

The level of background remaining in the $\pippim$ sample
is estimated by extrapolating an exponential fitted to the $\ptpsq$ distribution
in a control region,
$800<\ptpsq<2000$\,MeV$^2$/$c^2$,
dominated by background,
into the signal region
$\ptpsq<200$\,MeV$^2$/$c^2$.
Due to limited statistics,
the lifetime bin width was increased from $0.1\tauS$ to $0.5\tauS$ for this procedure.
The fraction of background in the $\KLpm$ sample is found to be
about $2\E{-3}$,
and no significant dependence of the background fraction on lifetime is observed in any bin of energy.
An appreciable lifetime dependence is not expected in this case
since both the signal and background are observed in the detector as two-track final states
with configurations which vary in similar fashion with the decay vertex longitudinal position.

\subsection{${\mathbf{K_S}}$ tagging}                   \label{sec:selection_tagging}

For $\pippim$ decays,
the good resolution on the transverse position of the decay vertex
allows a clean separation of decays from the $\KS$ and $\KL$ beams;\,
decays with a vertex position more than 4\,cm above the $\KL$ beam axis
after extrapolation of the reconstructed parent kaon trajectory back to
the position of the AKS counter
were classed as belonging to the $\KS$ beam (``vertex tagging'').
For $\pizpiz$ decays,
only the longitudinal position of the decay vertex can be reconstructed.
In this case, the identification of decays from the $\KS$ beam is accomplished using a
tagging station (Tagger)
traversed by the proton beam during transport to the $\KS$ target
and consisting of two scintillator ladders (one horizontal, one vertical)~\cite{bib:tagger}.
Each scintillator counter has a time resolution of $\sim$140\,ps,
and a proton crosses at least two counters.
A $\pizpiz$ decay is classed as belonging to the $\KS$ beam
if a signal is observed in the Tagger within $\pm2$\,ns of the reconstructed
event time.
A small fraction, $\alphaSL^{00}=(1.6\pm0.5)\E{-4}$, of $\KSzz$ decays
are mistagged as belonging to the $\KL$ beam
due to Tagger detection inefficiencies.
A larger fraction, $\alphaLS^{00}=(10.692\pm0.020)$\%, of $\KLzz$ decays
are mistagged as belonging to the $\KS$ beam
due to accidental proton signals in the Tagger in time with the event.
The values of $\alphaSL^{00}$ and $\alphaLS^{00}$
were inferred from studies of the mistagging fractions $\alphaSL^{+-}$ and $\alphaLS^{+-}$
for vertex tagged $\pippim$ decays,
combined with studies of $\KLzz$ and $\KLzzz$ decays
to determine the small differences in mistagging rates between charged and neutral modes.

\subsection{Event samples}

Applying the selection cuts above to the data recorded by NA48 during 1998 and 1999
yielded samples of
13.2M $\KSpm$, 12.2M $\KLpm$, 3.1M $\KSzz$ and 2.8M $\KLzz$ events
with reconstructed lifetimes in the range $0.5<\ctau/\tauS<3.5$,
where the number of $\pizpiz$ events has been corrected for $\KS$-$\KL$ mistagging.

Samples of simulated $\Kzpipi$ events corresponding to about 3-5 times the
data statistics were also available for analysis.
The Monte Carlo included a detailed modelling of the $\KS$ and $\KL$ beams
(including the $\KS$ beam halo)
and used the GEANT package~\cite{bib:geant} for particle tracking and the simulation of
processes such as multiple scattering, photon conversion, and secondary interactions.
Simulation of the LKr calorimeter response was based on a
library of electromagnetic and hadronic showers generated using GEANT.
The simulated events were passed through the same reconstruction
and selection code as the data.
Asymmetric non-Gaussian tails in photon shower energies 
due to hadron photoproduction in the liquid krypton,
which arise in about $3\E{-3}$ of cases,
were modelled using a parameterisation which was applied
to the reconstructed shower energies in the Monte Carlo events.

\subsection{Acceptance correction}

Since the $\KS$ and $\KL$ beams are almost collinear,
and since $\KS$ and $\KL$ decays are recorded simultaneously
using a common trigger,
the acceptances for $\KS$ and $\KL$ decays as a function of energy and position
are equal to good approximation,
and the detector acceptance essentially cancels
in the $\KS/\KL$ ratio of lifetime distributions from the two beams.
For the $\pippim$ decay mode,
effects due to decay in flight of the charged pions also cancel
in the $\KS/\KL$ ratio.

Small acceptance differences arise
due to the different divergences and transverse profiles of the $\KS$ and $\KL$ beams
and the consequent differences in illumination of the detector.
Also, the different definition of the upstream edge of the decay region for the two beams
introduces large acceptance differences at low values of the reconstructed lifetime.
The AKS veto requirement effectively represents a cut $\ctau>0$
on the true lifetime of decays in the $\KS$ beam
while the upstream acceptance for $\KL$ decays 
extends smoothly into the region $\ctau<0$.
This difference is clearly seen in Figure~\ref{fig:fit_ctau}
where examples of the uncorrected reconstructed lifetime distributions
for the $\KS$ and $\KL$ beams are plotted.

Acceptance differences for $\KS$ and $\KL$ decays were accounted for
by dividing the $\KS/\KL$ ratio observed in the data
by the acceptance ratio predicted by the Monte Carlo.
The acceptance correction also took
detector resolution effects (smearing) into account by defining
the $\KS$ and $\KL$ acceptances as the ratio
of the number of simulated events passing the selection cuts
in a bin of {\em reconstructed} energy and lifetime
to the number of generated events
in the corresponding bin defined using the {\em true} energy and lifetime.

Resolution induced acceptance differences due to the AKS veto requirement are largely eliminated
by restricting the lifetime range considered to the region $\ctau>0.5\tauS$.
A small (up to $\sim2$\%) residual variation of the acceptance correction
remains for $\pizpiz$ events at low energy and lifetime,
but the correction for $\pizpiz$ events is otherwise independent of lifetime
within the errors due to Monte Carlo statistics.
For $\pippim$ events,
the $\KS/\KL$ acceptance ratio is independent of lifetime at lower energies,
but falls significantly (by up to $\sim7$\%) with increasing lifetime at higher energies.
This is not due to an intrinsic difference between the $\KS$ and $\KL$ acceptances,
but rather to a slow decrease of the acceptance
with increasing radial distance of the decay vertex from the beam axis,
convoluted with the larger transverse size of the $\KS$ beam.
The fall in acceptance with increasing radius
is due dominantly to the geometrical cuts
and to the cut on the track momentum asymmetry,
and can be reliably modelled and checked by comparing the relevant reconstructed
distribution in data and Monte Carlo.
Similarly,
the modelling of the transverse beam profiles is checked
by comparing the reconstructed $\rcog$ distributions in data and Monte Carlo.

\section{Results}                                          \label{sec:results}

\begin{figure}[p]
\begin{center}
  \epsfig{file=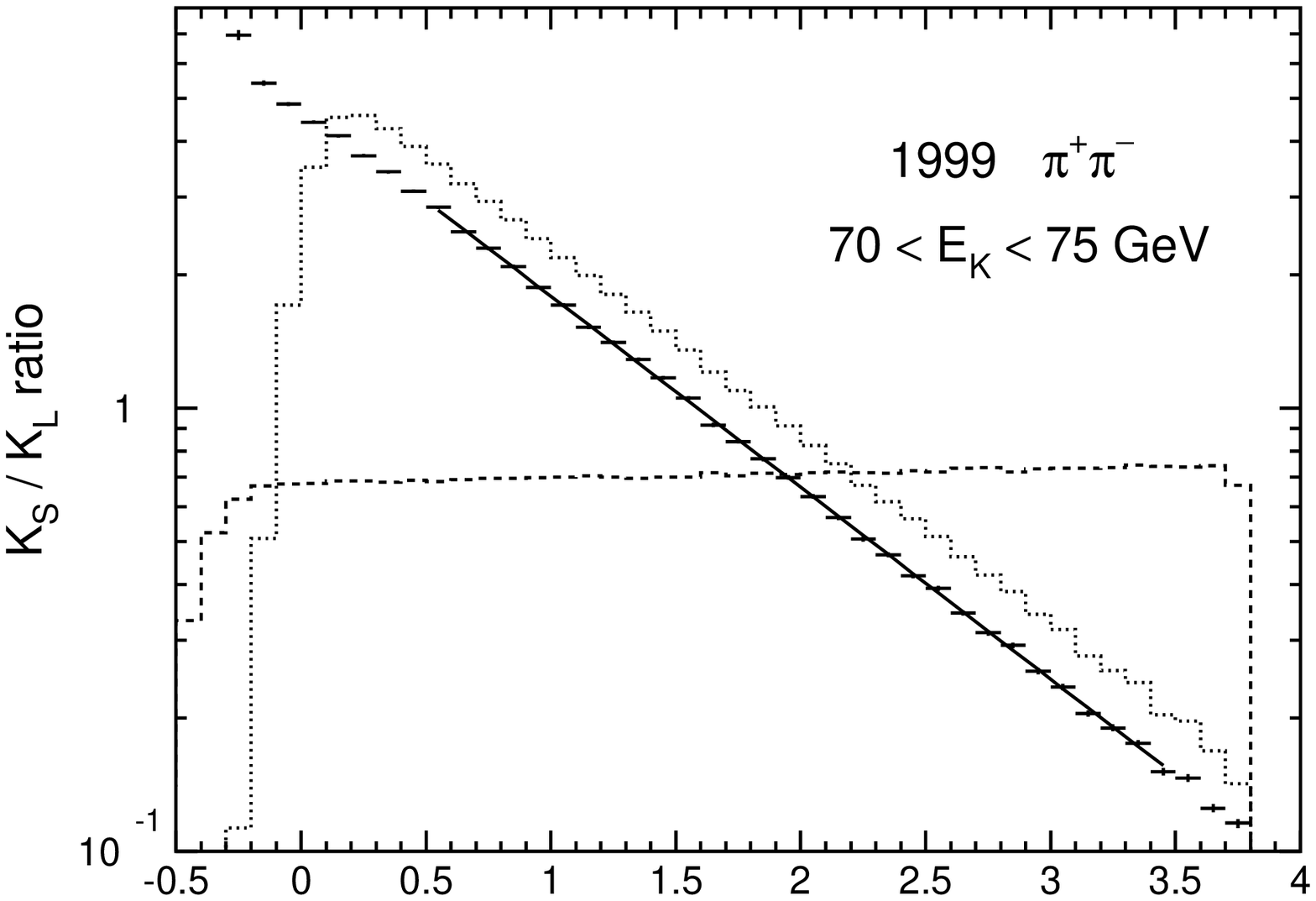,width=75mm}
  \epsfig{file=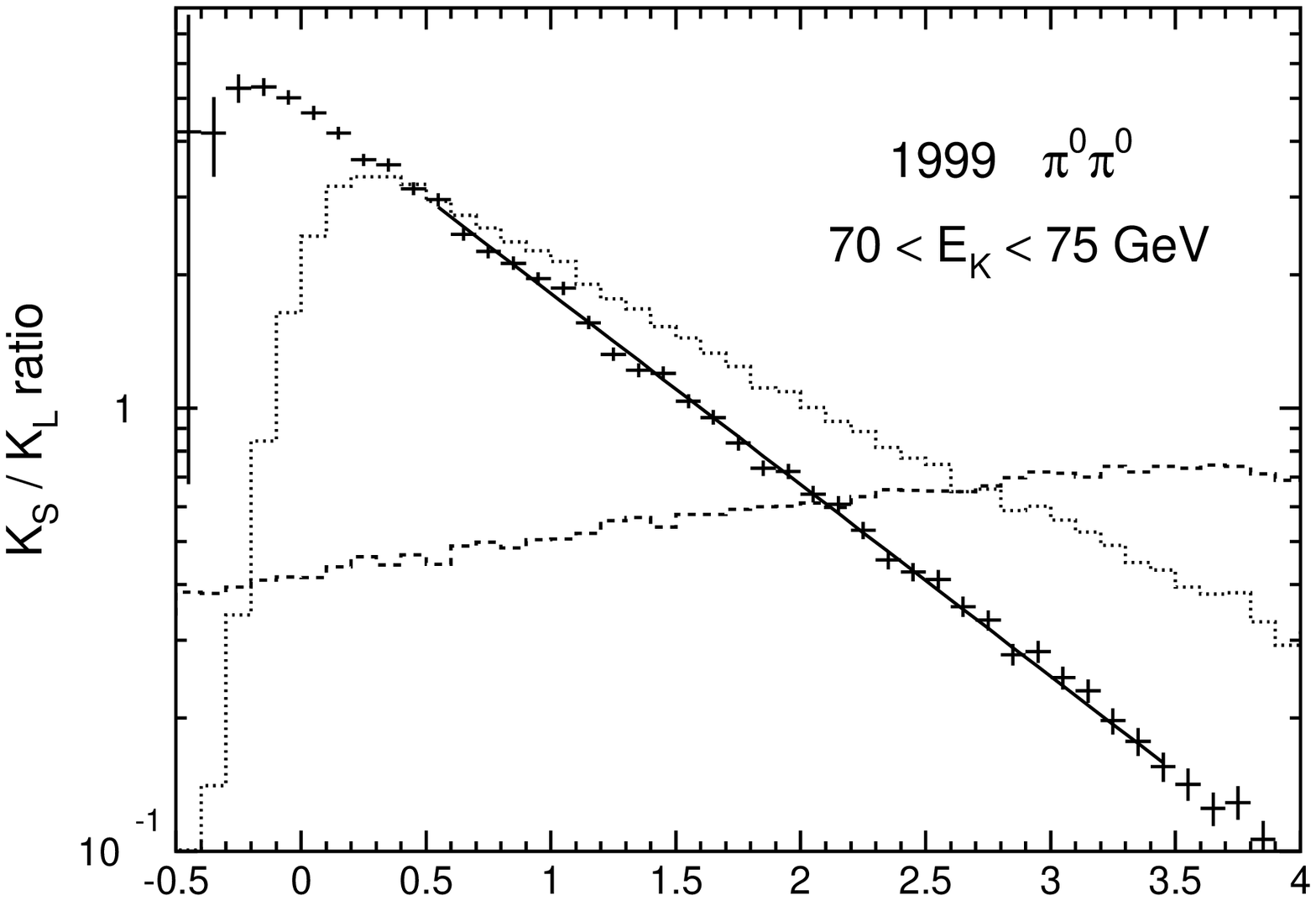,width=75mm}
  \epsfig{file=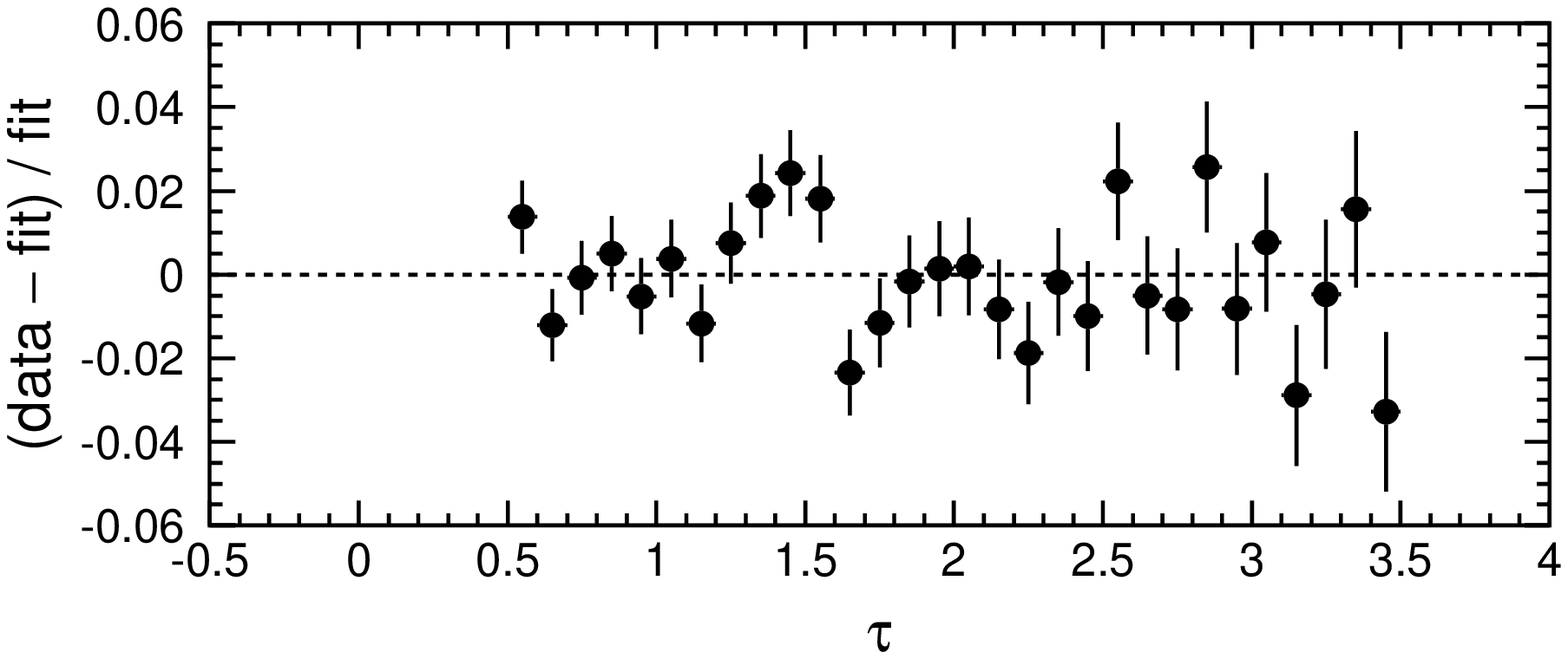,width=75mm}
  \epsfig{file=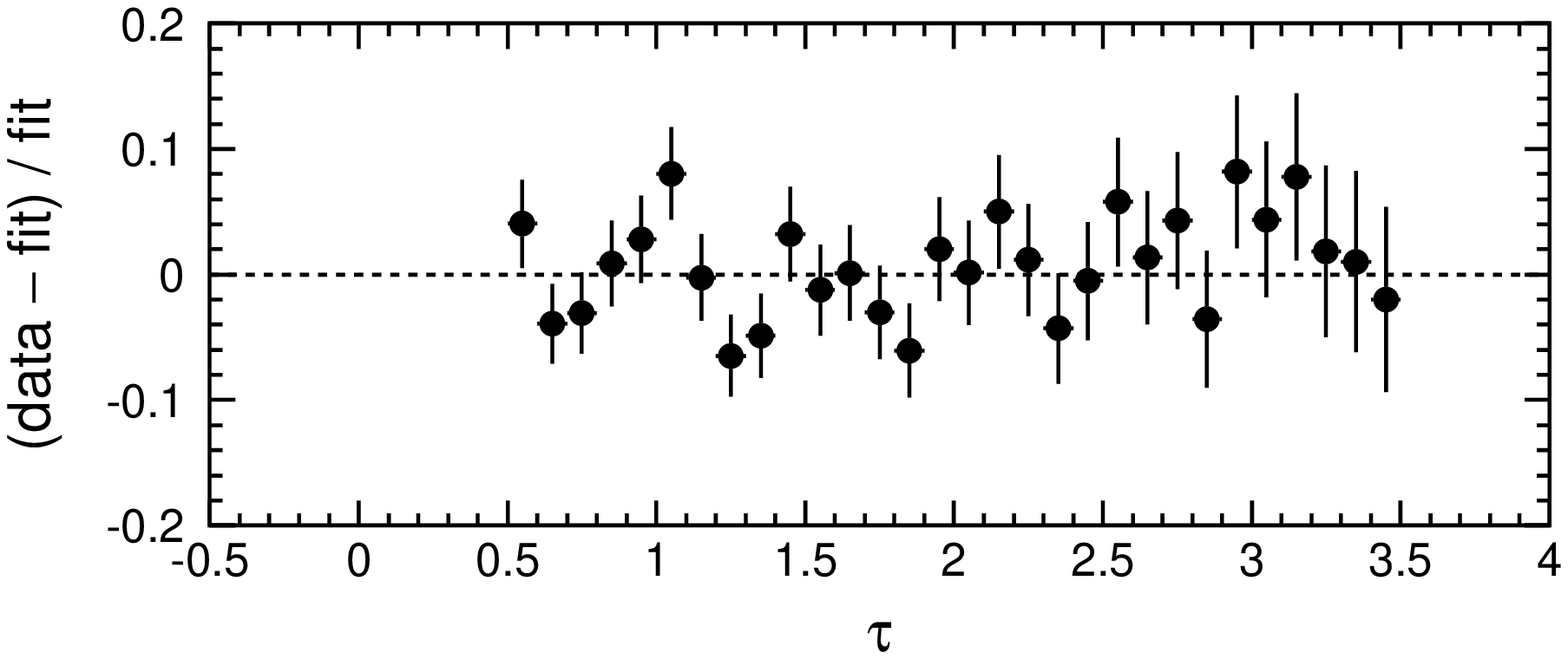,width=75mm}
  \\ ~~ \\
  \epsfig{file=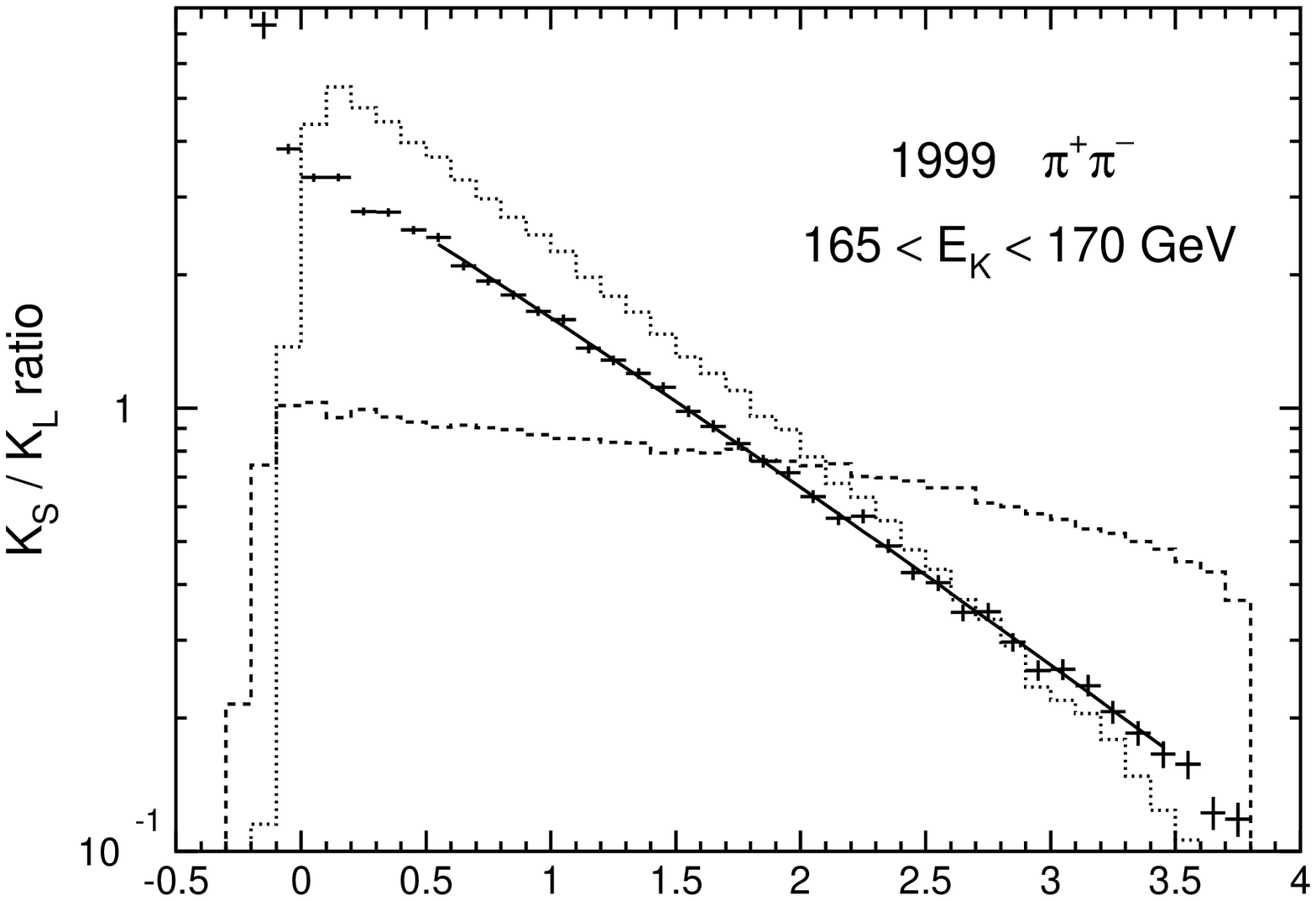,width=75mm}
  \epsfig{file=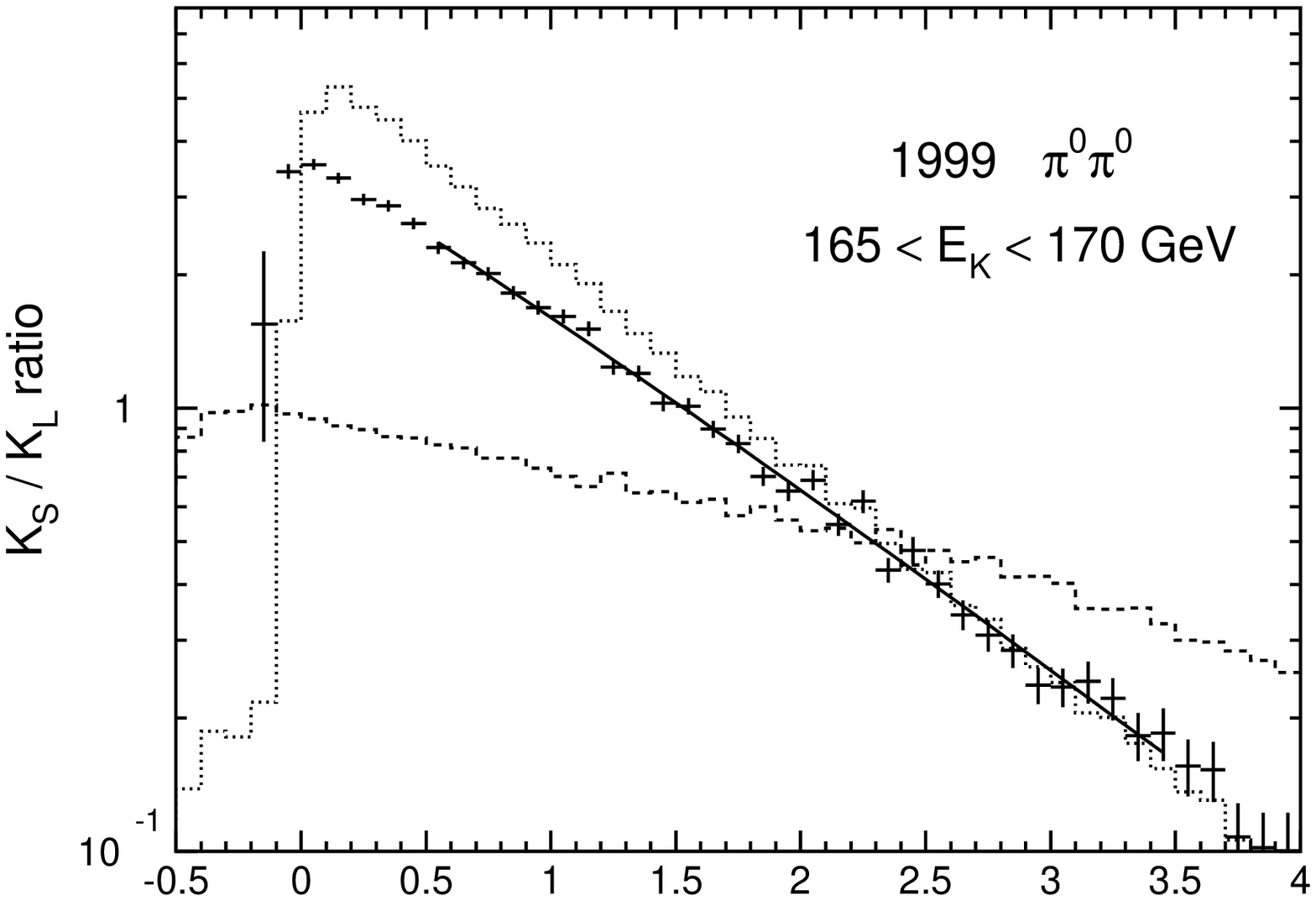,width=75mm}
  \epsfig{file=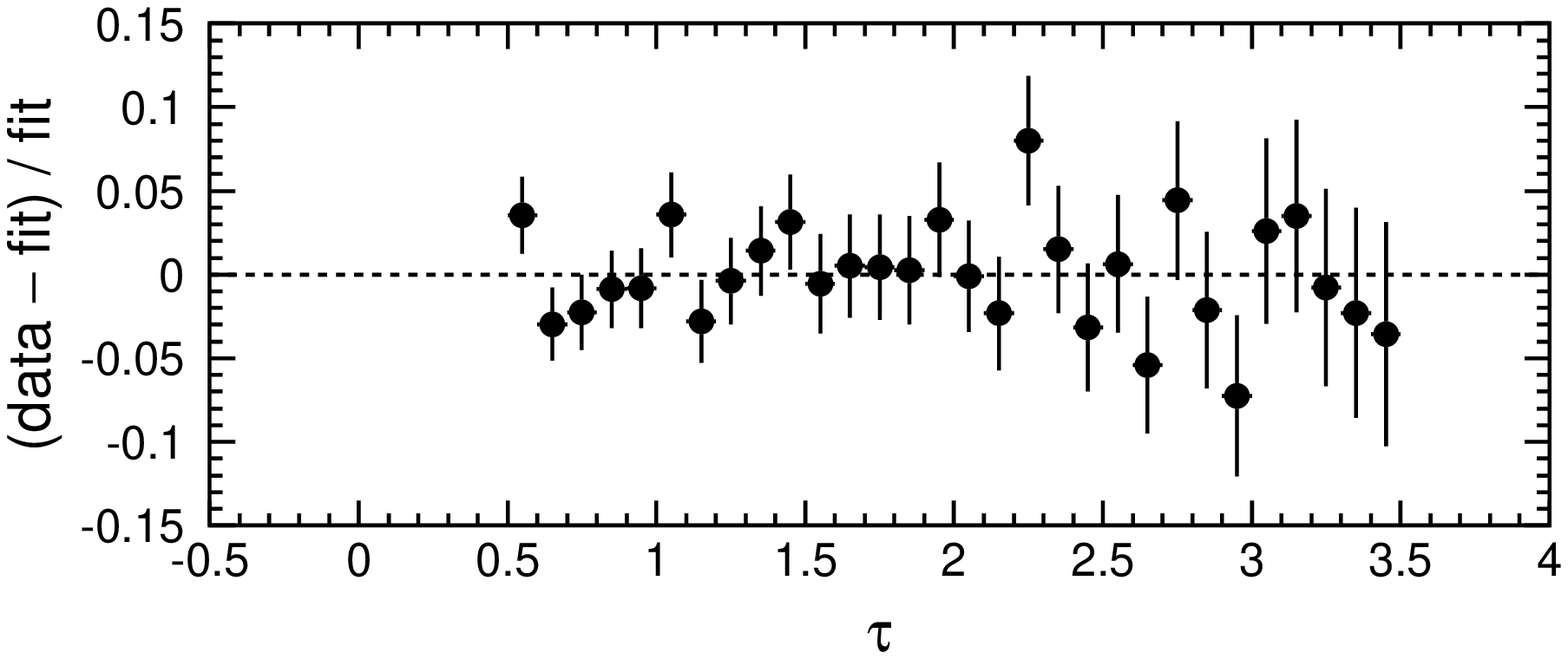,width=75mm}
  \epsfig{file=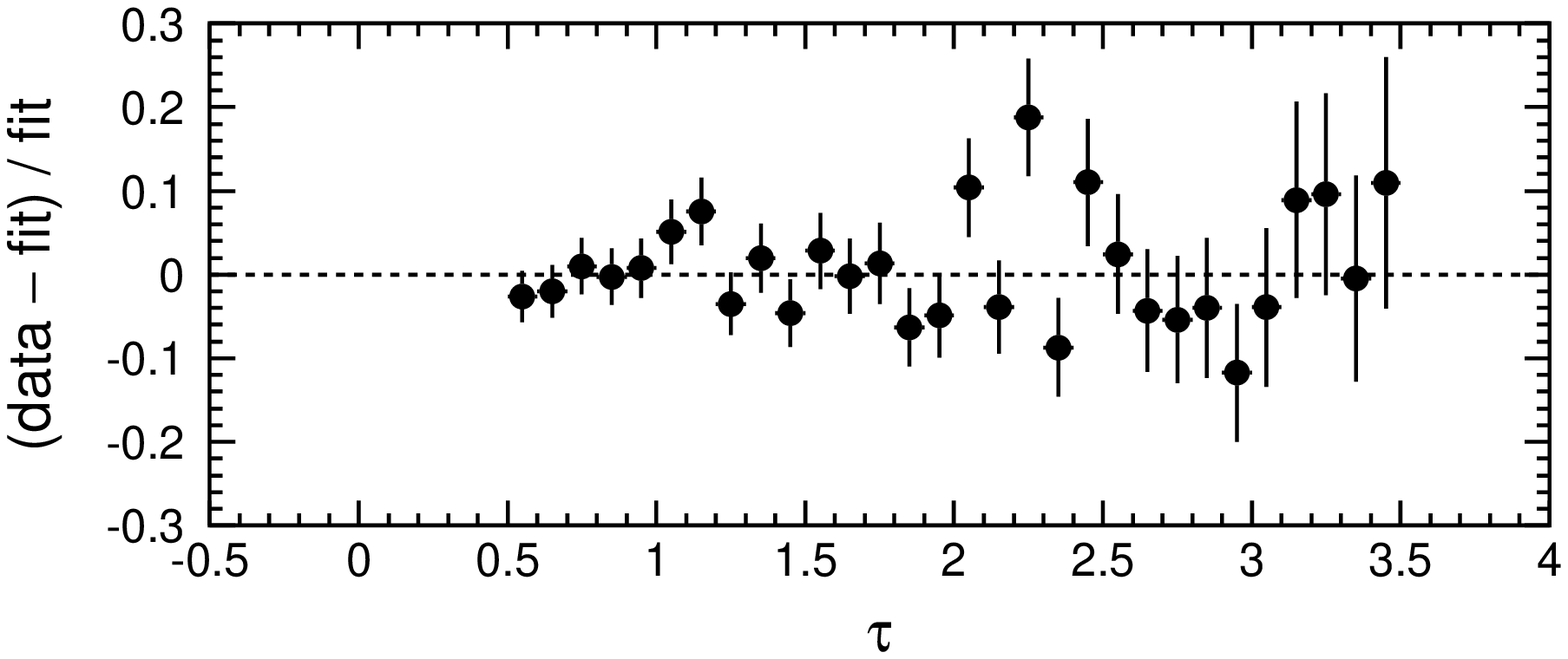,width=75mm}
  \caption{The points with error bars in the upper plot of each pair
           show examples of the corrected $\KS/\KL$ ratio as a function of
           the reconstructed proper lifetime, $\ctau$,
           expressed in nominal $\KS$ lifetime units of $0.8927\E{-10}$\,s.
           The dotted and dashed histograms show the corresponding uncorrected lifetime
           distributions for the $\KS$ and $\KL$ beams, arbitrarily normalised.
           The curves show the results of the fit for the $\KS$ lifetime,
           while the lower plot of each pair shows the normalised fit residuals.}
  \label{fig:fit_ctau}
\end{center}
\end{figure}

The $\KS/\KL$ ratio $R=\NS/\NL$ 
is computed in two-dimensional bins of energy and lifetime
of width $\Delta\EK=5$\,GeV and $\Delta\ctau=0.1\tauS$,
and corrected bin-by-bin for residual background in the $\KLpm$ and $\KLzz$ samples
and for small differences in the $\KS$ and $\KL$ acceptances, as described above.
For the $\pizpiz$ mode, the $\KS/\KL$ ratio is also corrected for $\KS$-$\KL$ mistagging.
Examples of the dependence of the corrected $\KS/\KL$ ratio
on the reconstructed proper lifetime $\ctau$
are shown in Figure~\ref{fig:fit_ctau}.
The shapes of the separate $\KS$ and $\KL$ lifetime distributions change appreciably with energy,
due largely to changes in the lifetime dependences of the $\pippim$ and $\pizpiz$ acceptances,
but the corrected $\KS/\KL$ ratio is approximately independent of energy.

The $\KS$ lifetime is determined by fitting a function of the form given in
Equations~(\ref{eqn:ratio}) and (\ref{eqn:ft})
to the $\KS/\KL$ ratio
over the lifetime range $0.5<\ctau/\tauS<3.5$ and energy range $70<\EK<170$\,GeV.
The following $\chi^2$ quantity is minimised in the fit:
\begin{equation*}
   \chi^2 = \sum_{i=1}^{20} \sum_{j=1}^{30}
       \left( R_{ij} - R(E_i,\ctau_j) \over \sigma_{ij} \right)^2
\end{equation*}
where $E_i$ is the central energy of the $i$'th bin of energy,
$\ctau_j$ is the central value of the $j$'th bin of proper lifetime,
$R_{ij}$ is the corrected $\KS/\KL$ ratio from the data,
$\sigma_{ij}$ is the statistical error on $R_{ij}$,
and $R(E_i,\ctau_j)=A_i \fS(\tS)/\fL(\tL)$
is the expected $\KS/\KL$ ratio.
The normalisation parameters $A_i=A(E_i)$ are allowed to vary in the fit.
At low energies, $70<\EK<140$\,GeV,
the values of the dilutions $\DS(\EK)$ and $\DL(\EK)$
appearing in the interference terms in the functions $\fS(\tS)$ and $\fL(\tL)$
are fixed using measurements of the dilution by the NA31 experiment~\cite{bib:na31_dilution}.
The dilutions $\DS(\EK)$ and $\DL(\EK)$ rise from about 0.2 at $\EK=70$\,GeV
to about 0.35 at $\EK=140$\,GeV and differ slightly
because of the different kaon production angles (4.2\,mrad and 2.4\,mrad)
used to define the $\KS$ and $\KL$ beams;\,
small corrections are applied to the NA31 measurements to account for the different
production angles used in this experiment.
At high energies, $140<\EK<170$\,GeV,
where the interference term contributes significantly to the ratio $R(\EK,\ctau)$,
the values $D_i=\DS(E_i)=\DL(E_i)$ of the dilution are allowed to vary in the fit,
neglecting the small differences between the $\KS$ and $\KL$ dilutions.
The remaining physics parameters appearing in the functions $f(t)$
are taken from the PDG averages of existing published measurements,
except the $\KS$ lifetime $\tauS$ which is a free parameter in the fit.

The fit was carried out separately for the $\pippim$ and $\pizpiz$ samples
from each year of data taking.
The values of the $\KS$ lifetime from each fit
are summarised in Table~{\ref{tab:fit_results}}.
The errors are statistical only and
include the contribution from finite Monte Carlo statistics.
The minimum value of the fit $\chi^2$ is also given in Table~{\ref{tab:fit_results}}
and corresponds in all cases to an acceptable fit probability.
Examples of the fit results and the fit quality
for the $\pippim$ and $\pizpiz$ data samples from the 1999 run
are shown in Figure~\ref{fig:fit_ctau}.
The effect of the interference terms in the fit function
is visible in the plots for the high energy bin as a slight curvature of the lines
representing the fit result.

\begin{table}
\begin{center}
\begin{tabular}{|c|c|c|c|c|}
\hline
   &   \multicolumn{2}{c|}{$\pippim$}  &   \multicolumn{2}{c|}{$\pizpiz$}  \\
\cline{2-5}
          &    $\tauS/10^{-10}$\,s     &  $\chi^2$/dof &   $\tauS/10^{-10}$\,s    &  $\chi^2$/dof  \\  
\hline
    1998  &   $0.89578 \pm 0.00109$    &  628.2/573   &   $0.89606 \pm 0.00247$   &  551.0/573   \\
    1999  &   $0.89598 \pm 0.00072$    &  601.2/573   &   $0.89635 \pm 0.00167$   &  543.5/573   \\
\hline
\end{tabular}
\begin{minipage}{0.9\linewidth}
\caption{Values of the $\KS$ mean lifetime
         and the minimum chi-squared per degree of freedom
         from the lifetime fits.
         The errors are statistical only.}
\label{tab:fit_results}
\end{minipage}
\end{center}
\end{table}

Consistent values of the fitted $\KS$ lifetime $\tauS$,
and of the fitted normalisation and dilution parameters $A_i$ and $D_i$,
were found for the charged and neutral modes
and for the samples from the different years of data taking.
The fitted values of the dilution $D_i$
within the energy range $140<\EK<170$\,GeV
were also consistent with the published NA31 measurements.

\section{Systematic errors}                \label{sec:systematics}

Various sources of systematic error on the measured $\KS$ lifetime were considered,
including uncertainties in the reconstruction of the energy and distance scales
for the charged and neutral decay modes,
and uncertainties due to the background subtraction, the tagging correction,
the Monte Carlo acceptance correction,
and the external physics parameters used in the fitting function.

As noted in Section~\ref{sec:selection_charged},
the energy and distance scales for reconstructed $\pippim$ events
are determined largely by the detector geometry,
especially of DCH1 and DCH2.
The systematic error due to uncertainties in detector geometry
was estimated by considering a variation of $\pm 2$\,mm in
the longitudinal separation of DCH1 and DCH2
and a variation of $\pm20$\,$\mu$m/m
in the relative transverse scale of the two chambers.
The position of the upstream edge of the $\KSpm$ decay vertex distribution
was found to be consistent within these tolerances
with the nominal position of the AKS counter.
In addition,
the momentum scale for reconstructed tracks was varied by $\pm0.1$\%,
as determined from the consistency of the reconstructed $\pippim$ invariant mass
with the kaon mass, $\mK$.
In each case,
the kaon decay reconstruction was redone for all events in the data
using the modified track positions or momenta,
and the lifetime analysis repeated.
The resulting change in the fitted $\KS$ lifetime
was taken as the estimate of the corresponding systematic error.

The systematic error due to uncertainties in the reconstruction of $\pizpiz$ events
was estimated by varying the energy scale of reconstructed showers by $\pm0.03$\%
and by varying the linearity and uniformity of the calorimeter response
by modifying the reconstructed shower energy by an amount
$\Delta E=\alpha + \beta E^2 + \gamma r E$,
where $r$ is the radial distance of the shower from the central detector axis.
The allowed variation in the neutral energy scale is determined from a comparison of the position of the edge of the
$\KSzz$ decay vertex distribution with the nominal position of the AKS counter.
The allowed ranges of the constants $\alpha$, $\beta$ and $\gamma$
are determined from studies of $\Kethree$, $\Kzzz$ and $\KLzzz$ decays,
and of $\pizero\to\gamma\gamma$ and $\eta\to\gamma\gamma$ decays in special runs with $\piminus$ beams.

The uncertainty due to the $\KLpm$ background subtraction was assessed
by conservatively assigning a systematic error equal to
$\pm 100$\% of the change in the fitted $\KS$ lifetime when the background
subtraction was removed.
The systematic error associated with the $\KLzz$ background subtraction
was estimated by varying the extrapolation factor $\lamext$
within the range $\lamext=1.2\pm0.2$.

\begin{table}[t]
\begin{center}
\begin{tabular}{|l|c|c|c|c|}
\hline
  Source &  Variation  &   $\pippim$  &  $\pizpiz$  &  $\pi\pi$   \\
\hline
  DCH1 radial scale          &  $\pm 20$\,$\mu$m/m                      &  $\pm 1.8$  &             &  $\pm 1.5$   \\
  DCH1 $z$ position          &  $\pm 2$\,mm                             &  $\pm 0.2$  &             &  $\pm 0.1$   \\
  Momentum scale             &  $\pm 0.001$                             &  $\pm 0.3$  &             &  $\pm 0.3$   \\
  LKr Energy scale           &  $\pm 0.0003$                            &             &  $\pm 1.7$  &  $\pm 0.3$   \\
  Energy loss ($\alpha$)     &  $\pm 10$\,MeV                           &             &  $\pm 3.4$  &  $\pm 0.6$   \\
  Non-linearity ($\beta$)    &  $\pm 0.00002$\,GeV$^{-1}$               &             &  $\pm 1.6$  &  $\pm 0.3$   \\
  Non-uniformity ($\gamma$)  &  $\pm 0.00001$\,cm$^{-1}$                &             &  $\pm 1.0$  &  $\pm 0.2$   \\
  LKr radial scale           &  $\pm 0.0003$                            &             &  $\pm 3.0$  &  $\pm 0.5$   \\
  Charged background         &  $\pm 100$\,\%                           &  $\pm 1.4$  &             &  $\pm 1.2$   \\
  Neutral background         &  $\lamext = 1.2 \pm 0.2$                 &             &  $\pm 2.1$  &  $\pm 0.3$   \\
  Collimator scattering      &  $\pm 100$\,\%                           &             &  $\pm 3.2$  &  $\pm 0.5$   \\
  MC: $\KS$ beam $y$ posn    &  $\pm 2$\,mm                             &  $\pm 1.6$  &  $\pm 0.2$  &  $\pm 1.3$   \\
  MC: $\KL$ beam $y$ posn    &  $\pm 2$\,mm                             &  $\pm 1.4$  &  $\pm 0.2$  &  $\pm 1.2$   \\
  MC: $\KS$ beam halo        &  $\pm 100$\,\%                           &  $\pm 1.3$  &  $\pm 1.3$  &  $\pm 1.4$   \\
  MC: non-Gaussian tails     &  $\pm50$\,\%                             &             &  $\pm 4.4$  &  $\pm 0.7$   \\
  MC: statistics             &                                          &  $\pm 3.0$  &  $\pm 4.9$  &  $\pm 2.7$   \\
  $\tau_{\mathrm{L}}$        &  $5.17 \pm 0.04\E{-8}$\,s                &  $\pm 0.1$  &  $\pm 0.1$  &  $\pm 0.1$   \\
  $|\eta_{+-}|$, $|\eta_{00}|$ &  $2.276, 2.262 \pm 0.017\E{-3}$        &  $\pm 0.1$  &  $\pm 0.1$  &  $\pm 0.1$   \\
  $\phi_{+-}$, $\phi_{00}$   &  $43.3\pm0.5\degree$, $43.2\pm1.0\degree$  &  $\pm 0.8$  &  $\pm 0.4$  &  $\pm 0.8$   \\
  $\Delta m$                 &  $0.5300\pm0.0012\E{10}$\,s$^{-1}$       &  $\pm 0.5$  &  $\pm 0.7$  &  $\pm 0.5$   \\
  $\DE$                      &  $\pm 2\,\sigma$                         &  $\pm 0.3$  &  $\pm 1.0$  &  $\pm 0.4$   \\
  $\alphaLS^{00}$            &  $0.10692 \pm 0.00020$                   &             &  $\pm 2.1$  &  $\pm 0.4$   \\
  $\alphaSL^{00}$            &  $(1.6 \pm 0.5)\E{-4}$                   &             &  $\pm 0.4$  &  $\pm 0.1$   \\
  Fit method                 &                                          &  $\pm 2.7$  &  $\pm 2.7$  &  $\pm 2.7$   \\
\hline
  \multicolumn{2}{|c|}{Total systematic error}                          &  $\pm 5.4$  & $\pm 10.9$  &  $\pm 5.1$   \\
\hline
  \multicolumn{2}{|c|}{Statistical error}                               &  $\pm 5.2$  & $\pm 12.9$  &  $\pm 4.8$   \\
\hline
\end{tabular}
\end{center}
\caption{Summary of systematic errors on the measured $\KS$ lifetime,
         in units of $10^{-14}$\,s.
         The final column corresponds to a combination of the $\pippim$ and $\pizpiz$ results.}
\label{tab:systematics}
\end{table}

The $\KLpm$ and $\KLzz$ samples contain a small fraction of events
arising from the genuine $\pi\pi$ decay
of neutral kaons produced by scattering in (dominantly) the final collimator
of the $\KL$ beam.
In the charged mode,
collimator scattered events are largely removed by the cut on $\ptpsq$
and contribute a fraction less than $10^{-4}$ to the $\KLpm$ sample.
This contribution is already included
in the charged background estimate described in Section~\ref{sec:selection_charged},
and no additional systematic error is warranted.
Studies of events with large transverse momentum ($\ptpsq>200$\,MeV$^2$/$c^2$)
showed that the lifetime distribution of collimator scattered events corresponds approximately
to an exponential decay characterised by the $\KS$, rather than $\KL$, lifetime,
and that their energy distribution is similar to that of unscattered $\KL$ decays.
The high $\ptpsq$ events were also used to estimate the fraction,
$(4.2\pm1.0)\E{-4}$~\cite{bib:collscat},
of collimator scattered events in the $\KLzz$ sample,
where no transverse momentum cut is possible.
For the $\pizpiz$ mode,
the effect of collimator scattering on the $\KS$ lifetime measurement
was studied by modifying the fit function to be of the form
$R(E_i,\ctau_j)=A_i \fS(\tS)/[\fL(\tL)+\Acoll e^{-\tcoll/\tauS)}]$
where $\tcoll=(z-\zcoll)(\mK/\pK)$ is the proper lifetime of the decay
relative to the position of the final $\KL$ collimator
and $\Acoll$ is a constant which is adjusted to give
the above fraction of collimator scattered events in the $\KLzz$ sample.
This fraction was assumed to be independent of energy.
The resulting systematic error
was conservatively estimated as $\pm 100$\% of the change in the $\KS$ lifetime
due to inclusion of the collimator scattering component in the fit.

The systematic error associated with uncertainties in the acceptance correction
was estimated by varying the assumed vertical position of the $\KS$ and $\KL$ beams
in the Monte Carlo within a range $\pm 2$\,mm around their nominal positions.
In addition,
the sensitivity to the transverse beam profile was studied by
switching off the simulation of the $\KS$ beam halo in the Monte Carlo.
In each case,
the resulting change in the fitted $\KS$ lifetime was taken as the systematic error.
The sensitivity of the acceptance correction to the modelling of the detector resolution
was assessed by removing the simulation of non-Gaussian tails
from the reconstructed photon shower energies,
and a systematic error of $\pm 50$\% of the effect on the fitted $\KS$ lifetime was assigned.
The effect of possible non-Gaussian tails in the reconstructed drift chamber hit positions
was also studied in the Monte Carlo,
but found to have a negligible effect on the fitted $\KS$ lifetime.
Finally, the component of the overall statistical error on the fitted lifetime
arising from the finite Monte Carlo statistics
was extracted, and classified as a separate systematic error.

The external physics parameters
$\tauL$, $\eta_{+-}$, $\eta_{00}$, $\phi_{+-}$, $\phi_{00}$ and $\Delta m$
appearing in the functions $f(t)$
were each varied in turn within a range given by the error on the PDG average value.
The mistagging fractions $\alphaLS^{00}$ and $\alphaSL^{00}$ were
varied within the uncertainties given in Section~\ref{sec:selection_tagging}.
The value of the dilutions $\DSL(\EK)$ in the energy range $70<\EK<140$\,GeV
was varied by twice the error on the NA31 measurements~\cite{bib:na31_dilution}.
The extra factor of two conservatively takes into account
uncertainties in correcting the NA31 measurements to the NA48 experiment.

The fitting procedure itself was tested
by applying the fit to the generated inclusive energy and lifetime distributions
of parent kaons from the Monte Carlo samples.
No significant bias was observed on the fitted $\KS$ lifetime,
and the statistical precision of the test was assigned as a systematic error.

The systematic errors on the $\KS$ lifetime from each of the above sources
for the 1998 and 1999 data combined
are summarised in Table~{\ref{tab:systematics}}.
For comparison,
the effect of the background subtraction is to increase the fitted $\KS$ lifetime
for the $\pizpiz$ mode by $12.8\E{-14}$\,s,
while the acceptance correction from the Monte Carlo changes the $\KS$ lifetime
by $+22.5\E{-14}$\,s and $-9.1\E{-14}$\,s
for the $\pippim$ and $\pizpiz$ modes, respectively.

The analyses carried out for each year of data taking and for each decay mode
are combined taking into account any correlations between the separate analyses.
The total systematic error is obtained by summing the individual errors in quadrature.
The measured values of the $\KS$ lifetime for each decay mode
and for the two decay modes combined are:
\begin{alignat*}{2}
   \tauS & = (0.89592 \pm 0.00052 \pm 0.00054) \E{-10}\,\mathrm{s}  & \qquad (&\pippim)  \\
   \tauS & = (0.89626 \pm 0.00129 \pm 0.00100) \E{-10}\,\mathrm{s}  & \qquad (&\pizpiz)  \\
   \tauS & = (0.89598 \pm 0.00048 \pm 0.00051) \E{-10}\,\mathrm{s}  & \qquad (&\pippim+\pizpiz)
\end{alignat*}
where the first error is statistical and the second systematic.

Various cross-checks were performed to verify the integrity and stability of the result.
No significant dependence of the fitted $\KS$ lifetime was found on
the lifetime or energy range used in the fit,
or on the energy range within which $\DE$ was allowed to vary in the fit.
The effect of varying the main selection cuts used in the analysis was also studied,
namely the cuts on
$\ptpsq$, $\rcog$, the momentum asymmetry $|p_+-p_-|/(p_++p_-)$,
and the minimum radii of tracks and clusters in DCH1 and the LKr calorimeter.
In each case,
either no statistically significant variation in the fitted lifetime was found,
or the observed variation was found to be within the systematic errors assigned.
The stability of the analysis was also tested by
dividing the $\pippim$ data sample into topologies for which
the positive and negative tracks initially curve towards or away from each other
in the spectrometer magnetic field;\,
consistent values of the fitted lifetime were found for the two topologies.
Similarly, selecting either of the two polarities of the spectrometer magnetic field setting
gave consistent results.
Other tests involved dividing the data samples according to
the primary data-taking periods,
the event time within the SPS spill,
and the azimuthal orientation of the decay.
For the charged mode,
separating $\KS$ and $\KL$ decays using the Tagger
in place of vertex tagging gave no significant change in the result.

\section{Summary}

The $\KS$ lifetime has been measured using $\KSpipi$ and $\KLpipi$ decays
recorded by the NA48 experiment in 1998 and 1999.
The combined result for the $\pippim$ and $\pizpiz$ decay modes,
$\tauS = (0.89598 \pm 0.00048 \pm 0.00051)\E{-10}$\,s,
has a precision better than that of
the current PDG average~\cite{bib:pdg} of existing measurements,
and lies about 1.7 standard deviations above it.


\begin{thebibliography}{99}
\bibitem{bib:collscat}
    The collimator scattering background fraction of $(9.6\pm2.0)\E{-4}$
    determined in ref.~\cite{bib:na48_eprime} refers to $\KL$ decays
    which have been weighted according to their reconstructed proper lifetime.
\end{thebibliography}
\end{document}